\documentclass[epj]{svjour}
\usepackage{graphicx}
\usepackage{amsmath}
\usepackage{amssymb}
\usepackage{color}
\usepackage[normalem]{ulem}
\begin{document}
\title{Effect of an oscillating Gaussian obstacle in a Dipolar Bose-Einstein condensate}
\author{S. Sabari$^{1,2}$ and R. Kishor Kumar$^{3}$}

\authorrunning{S Sabari}
\titlerunning{Oscillating obstacle in a dipolar BEC}

\institute{$^{1}$Department of Physics, Bharathidasan University, Palkalaiperur Campus, Tiruchirappalli, 620024, India \\ $^{2}$Department of Physics, Savitribai Phule Pune University, Pune 411007, Maharashtra, India \\ $^{3}$Instituto de F\'{i}sica, Universidade de S\~{a}o Paulo, 05508-090 S\~{a}o Paulo, Brazil}

%
\abstract{
We study the dynamics of vortex dipoles in erbium ($^{168}$Er) and dysprosium ($^{164}$Dy) dipolar Bose-Einstein condensates (BECs) by applying an oscillating blue-detuned laser (Gaussian obstacle). For observing vortex dipoles, we solve a nonlocal Gross-Pitaevskii (GP) equation in quasi two-dimensions in real-time. We calculate the critical velocity for the nucleation of vortex dipoles in dipolar BECs with respect to dipolar interaction strengths. We also show the dynamics of the group of vortex dipoles and rarefaction pulses in dipolar BECs. In the dipolar BECs with Gaussian obstacle, we observe rarefaction pulses due to interaction of dynamically migrating vortex dipoles.
} 

\maketitle

\section{Introduction}
\label{sec1}

The remarkable observation of Bose-Einstein condensates (BECs) in $^{52}$Cr~\cite{Lahaye:2007,Griesmaier:2006}, $^{164}$Dy~\cite{Lu:2011,Youn:2010}, and $^{168}$Er~\cite{Aikawa:2012,rev1} with both dipole-dipole interaction (DDI) and \textit{s-wave} contact interaction has opened a wholly new exciting field that continues to thrive~\cite{dbec1,dbec2,FetterRMP}. Contrary to the short-range contact interaction, the DDI is a long-range anisotropic interaction that can be either repulsive or attractive. The \textit{s-wave} contact interaction, $a_s$, is experimentally controllable by Feshbach resonance~\cite{FBR}. It is therefore appealing to study the properties of dipolar BECs in variable short-range contact interaction regimes. However, the DDI is also inherently controllable, either via the magnitude of the external electric field, or by modulating the external aligning field in time, which allows to tune the magnitude and sign of the DDI~\cite{tuneDDI}. Due to the long-range nature and anisotropic character of the DDI, the dipolar BEC possesses many distinct features and new phenomena such as the new dispersion relations of elementary excitations~\cite{Wilson:2010,Ticknor:2011}, unusual equilibrium shapes, the roton-maxon character of the excitation spectrum~\cite{Santos:2000,Yi:2003,Ronen:2007,Parker:2008}, quantum phases including supersolid and checkerboard phases~\cite{Tieleman:2011,Zhou:2010}, anisotropic solitons~\cite{equ2Db,solRK,solPM}, vortices~\cite{rev4,rev5}, hidden vortices~\cite{Sabari2017} and distinct vortex lattices including crater-like structure, square lattices~\cite{vor1,vor2,vor3}. 

The modern optical techniques help to control the parameters of the condensate and visualize topological defects such as rarefaction pulses and quantized vortices in BECs~\cite{FetterRMP,equ2Db,solRK,solPM,rev4,rev5,vor1,vor2,vor3}. Recently, vortex tangles caused by an oscillatory perturbation were observed experimentally. The vortex tangle configuration is a signature of the presence of a quantum turbulent regime in the BEC cloud \cite{Henn09}. Moreover, recent studies on quantum turbulence are still concentrating on understanding the dynamics of quantized vortices \cite{PLTP}. Vibrating structures such as spheres, grids, and wires are used in superfluid $^3$He and $^4$He to create quantum turbulence \cite{PLTP,Hanninen07}. Despite the differences between these structures, the experiments show surprisingly similar behavior.
 
Introducing an oscillating potential in atomic dipolar BECs will be helpful to analyze the intrinsic nucleation of topological defects and synergy dynamics of vortices and rarefaction pulses. Also, this technique suggests a powerfull method for making quantum turbulence in trapped dipolar BECs, in addition to the other methods that have been used so far~\cite{Henn09,Berloff02,Kobayashi07} in alkali BECs. Eventually, the dynamics of vortices and rarefaction pulses can be visualized in atomic dipolar BECs, which enabling experimental and theoretical challenges for further analysis.
Up to now, vortex dipoles caused by oscillating potentials in alkali BECs have been observed in experiments and compared to theoretical models~\cite{osc1,osc2,Jackson00,Raman99,Onofrio20,Neely10,rev2,rev3}. Nonlinear dynamical behaviors, critical velocity for vortex dipoles, hydrodynamic flow, vortices, rarefaction pulses and other interesting perspectives have been carried out in alkali BECs using the oscillating Gaussian potential~\cite{osc1,osc2,Jackson00,Raman99,Onofrio20,Neely10}.

Inspite of the many experiments that have been carried out on $^{164}$Dy and  $^{168}$Er condensates there has still been no experimental observation of vortices in dipolar BECs. So, investigating the dynamics of vortex dipoles in a dipolar BEC by introducing an oscillating potential will be a fascinating experimental exploration. Thus, this model will be helpful to perform new experiments with the aim of observing vortices in dipolar BECs. In the present work, we are interested in studying the nucleation and dynamics of vortex dipoles and rarefaction pulses. 

The next sections are organized as follows. In Sec.~\ref{sec:frame}, we present the general three-dimensional mean-field equation for the dipolar BECs and the corresponding two-dimensional (2D) reduction. In Sec.~\ref{sec:numerical}, we present our numerical results, where we include plots on the critical velocity for the nucleation and dynamics of vortex-dipoles. Further, in this section, we show the rarefaction pulses due to the annihilation of vortex-dipoles. Finally, in Sec.~\ref{sec:con}, we present a summary of our conclusions and perspectives.

\section{The mean-field formalism}
\label{sec:frame}
At ultralow temperatures, a dipolar BEC is described by the time-dependent GP equation with a nonlocal integral corresponding to the DDI~\cite{dbec1,dbec2,lasPM,rev5,rev6}

\begin{align}
 i\hbar\frac{\partial \phi({\mathbf r},t)}{\partial t}& =\Big(-\frac{\hbar^2}{2m}\nabla^2+V({\mathbf r},t) + g \left\vert \phi({\mathbf r},t)\right\vert^2 \Big)\phi({\mathbf r},t)\notag \\ &+N \int U_{\mathrm{dd}}({\mathbf  r}-{\mathbf r}')\left\vert\phi({\mathbf r}',t)\right\vert^2 d{\mathbf r}'\phi({\mathbf r},t), \label{eqn:dgpe}
\end{align}

\noindent with $({\bf r},t)=({\bf \rho},t)$ and the radial coordinate being $\rho=\sqrt{x^2+y^2}$. The trapping potential $V({\mathbf r},t) = V_{ext} + V_G$ contains a cylindrically symmetric harmonic trap in addition to a Blue detuned Gaussian obstacle. The cylindrically symmetric trap is $V_{ext}({\mathbf r})=\frac{1}{2} m (\omega_\rho^2 (x^2+y^2)+ \omega_z^2 z^2)$, with $\omega_x = \omega_y = \omega_\rho$ and $\omega_z$ being the radial and axial trap frequencies respectively. The trap aspect ratio of the harmonic trap is $\lambda=\omega_{z}/\omega_{\rho}$. The Gaussian obstacle is

\begin{equation}
V_{G}(\rho,t) = V_{0} \exp\left(-\frac{\left[x-x_0(t)\right]^2+y^2}{w_0^2}\right),
\end{equation}
where $V_0$, $x_0(t)$ and $w_0^2$ are the height, position and width of the Gaussian obstacle. The position of the obstacle $x_0(t)=\epsilon \sin(\omega t)$ provides parametric resonance with respect to the oscillating frequency $\omega$. One can control the velocity ($v=\epsilon\omega$) of oscillation of the obstacle with respect to the amplitude $\epsilon$ and the frequency $\omega$. In the present study, $\epsilon = 10 \mu$m and $\omega = 60 /s$. However, the velocity of the obstacle also depends on $V_{0}$ and $w_0$, and we keep these fixed: $V_0=80\,\hbar\omega_\rho$ and $w_{0}=0.25\,\mu m$.    
The two-body contact interaction strength is $g=4\pi$ $\hbar^2a_s N/m$ where $a_s$, $m$, and $N$ are atomic scattering length, mass of the atom and number of atoms respectively. We consider that the magnetic dipoles are polarized along $z$ direction and the corresponding dipolar interaction term is $ U_{\mathrm{dd}}(\bf R)=(\mu_0 \mu^2/4\pi)(1-3\cos^2 \theta/ \vert  {\bf R} \vert  ^3)$, where the relative position of the dipoles is ${\bf R= r -r'}$, $\theta$  is the angle between ${\bf R}$ and the direction of polarization $z$, $\mu_0$ is the permeability of free space and $\mu$ is the dipole moment of the condensate atom. In the present study, we consider the $^{168}$Er and $^{164}$Dy atoms, their corresponding dipole moments being $\mu=7\mu_B$ and $10\mu_B$ respectively. The normalization of the mean-field wavefunction is $\int d{\bf r}\vert\phi({\mathbf r},t)\vert ^2=1.$

It is convenient to use the GP equation (\ref{eqn:dgpe}) in a dimensionless form. For this purpose, we introduce the dimensionless variables ${\bar {\bf r}}= {\bf r}/l,{\bar {\bf R}}={\bf R}/l, \bar a_s=a_s/l, \bar a_{\mathrm{dd}}=a_{\mathrm{dd}}/l, \bar t=t\bar \omega_\rho$, $\bar x=x/l, \bar y=y/l, \bar z=z/l, \bar \phi=l^{3/2}\phi$, $l=\sqrt{\hbar/(m \omega_\rho)}$. Eq. (\ref{eqn:dgpe}) can be rewritten (after removing the overhead bar from all the variables) as 
\begin{align}
i \frac{\partial \phi({\mathbf r},t)}{\partial t} & = \Big(-\frac{1}{2}\nabla^2+V({\mathbf r},t) +g \vert {\phi({\mathbf{r}},{t})} \vert^2 \Big) \phi({\mathbf r},t) \\ \nonumber &+ 3N a_{\mathrm{dd}}\int \frac{1-3\cos^2\theta}{\vert \bf{R}\vert^3} \vert \phi({\mathbf{r}}',t) \vert^2 d{\mathbf{r}}' \phi({\mathbf r},t),\label{gpe3d} 
\end{align}

To compare the dipolar and contact interactions, often it is useful to introduce the length scale $a_{\mathrm{dd}}\equiv \mu_0 \mu^2 m/(12\pi \hbar^2 l)$ and its values for $^{164}$Er and $^{168}$Dy are $66a_0$ and $131a_0$ respectively, with $a_0$ being the Bohr radius~\cite{dbec1}.

The stability of dipolar BEC depends on the external trap geometry, \textit{e.g.} a dipolar BEC is stable or unstable depending on whether the trap is pancake- or cigar-shaped, respectively. The instability usually can be overcome by applying a strong pancake trap and applying repulsive two-body contact interaction. The external trap helps to stabilize the dipolar BEC by imprinting anisotropy to the density distribution. Hence, we carry out the present study with axially-symmetric pancake-shaped magnetic trap.

So, we assume that the dynamics of the dipolar BEC in the axial direction is strongly confined with the ground state $\phi(z)=\exp(-z^2 /2d_z^2)/(\pi d_z^2)^{1/4}, \quad d_z= \sqrt{1/\lambda},$ and the wave function \begin{align}\phi({\bf r})\equiv \phi(z) \psi(\rho,t)=(1/(\pi d_z^2)^{\frac{1}{4}})\exp(-z^2/2d_z^2) \psi(\rho,t),\nonumber \end{align} where axial harmonic oscillator length $d_z=\sqrt{1/(\lambda)}$  and  ${\bf\rho} \equiv (x,y)$.
Therefore, it is more suitable to consider the system in quasi two-dimensions. One can obtain the effective 2D equation for the pancake-shaped dipolar BEC by integrating the equation (\ref{gpe3d}) over the $z$ direction using the above wave function $\phi({\bf r})$~\cite{equ2Da,equ2Db,lasPM}
\begin{align}
 i\frac{\partial \psi(\rho,t) }{\partial t}& =\Big(-\frac{1}{2}\nabla_\rho^2+ V_{2D} + g_{2} \, \vert \psi(\rho,t) \vert^2 \Big) \psi(\rho,t) \\ \notag &+g_{d}\int \frac{d{\bf k}_\rho}{(2\pi)^2} e^{-i{\bf k}_\rho.{\bf\rho}}\widetilde n({\bf k}_\rho,t)h_{2D}\left(\frac{k_\rho d_z}{\sqrt{2}}\right) \psi(\rho,t).\label{gpef}
\end{align}
 where, two-body contact interaction $g_{2}= 4\pi Na_s/(\sqrt{2\pi}d_z)$, dipolar interaction $g_{d}=4\pi Na_{\mathrm{dd}}/(\sqrt{2\pi}d_z)$, and $k_\rho=(k_x^2+k_y^2)^{1/2}$. The dipolar term has been written in the Fourier space~\cite{equ1,equ2}.
\section{Dynamics of vortices}
\label{sec:numerical}

The results in this section were obtained within a full numerical approach to solve 2D nonlinear differential Eq.~(4). When we have dipolar interactions, we have to combine the split-step Crank-Nicholson method and Fast Fourier Transform (FFT) for evaluating dipolar integrals in momentum space as in Ref.~\cite{equ1,equ2,CUDA}. 
\begin{figure*}[!ht]
\begin{center}
\includegraphics[width=14.0cm]{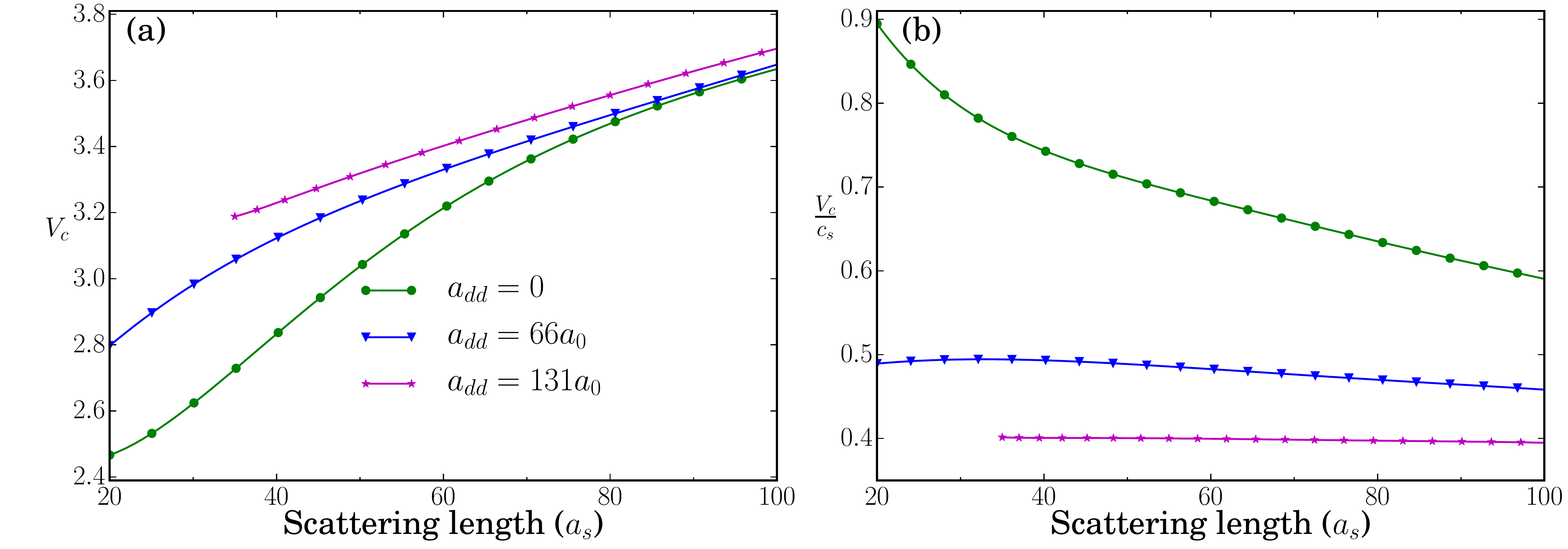}
\end{center}
\caption{(color online) (a) Plot of the critical velocity  ($V_c$) Vs s-wave scattering length ($a_s$) and (b) plot of the $V_c/c_s$ Vs $a_s$ for the three different cases, nondipolar (Green line with circles), $^{168}$Er (Blue line with triangles), and $^{164}$Dy (Magenta line with stars) condensates, respectively. The numerical simulations are carried out for  $N=2\times10^4$.}
\label{f1}
\end{figure*}
For looking at the dynamics of vortex dipoles, the 2D numerical simulations were carried out in real time with a grid size having $512$ points for each dimension, where we have $\Delta x = \Delta y = 0.05$ for the space-steps and $\Delta t = 0.00025$ for the time-step. Also, the results were verified by doubling the aforementioned grid sizes. 
\begin{figure}[!ht]
\begin{center}
\includegraphics[width=9cm,height=3.5cm]{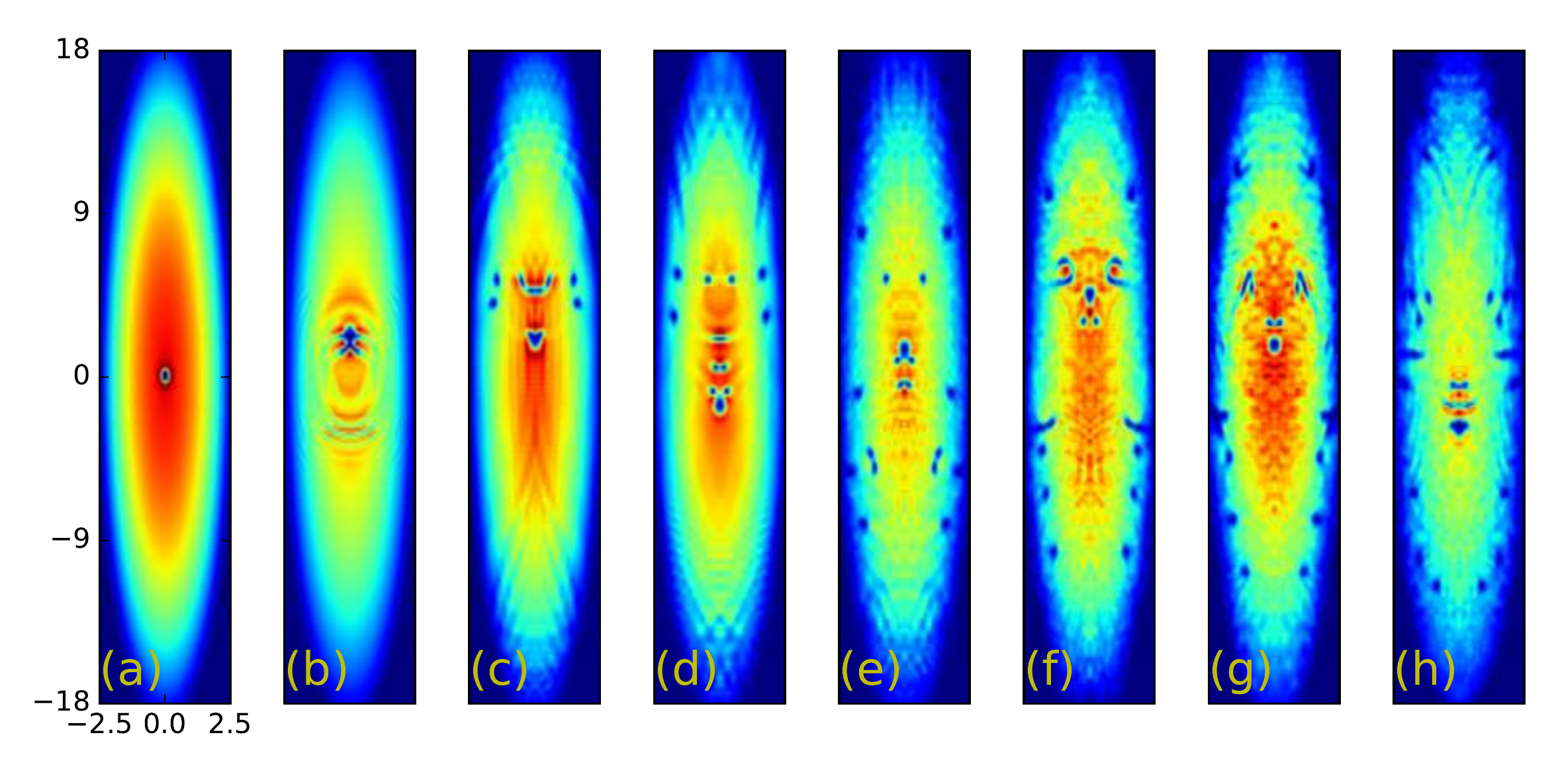}
\includegraphics[width=9cm,height=3.5cm]{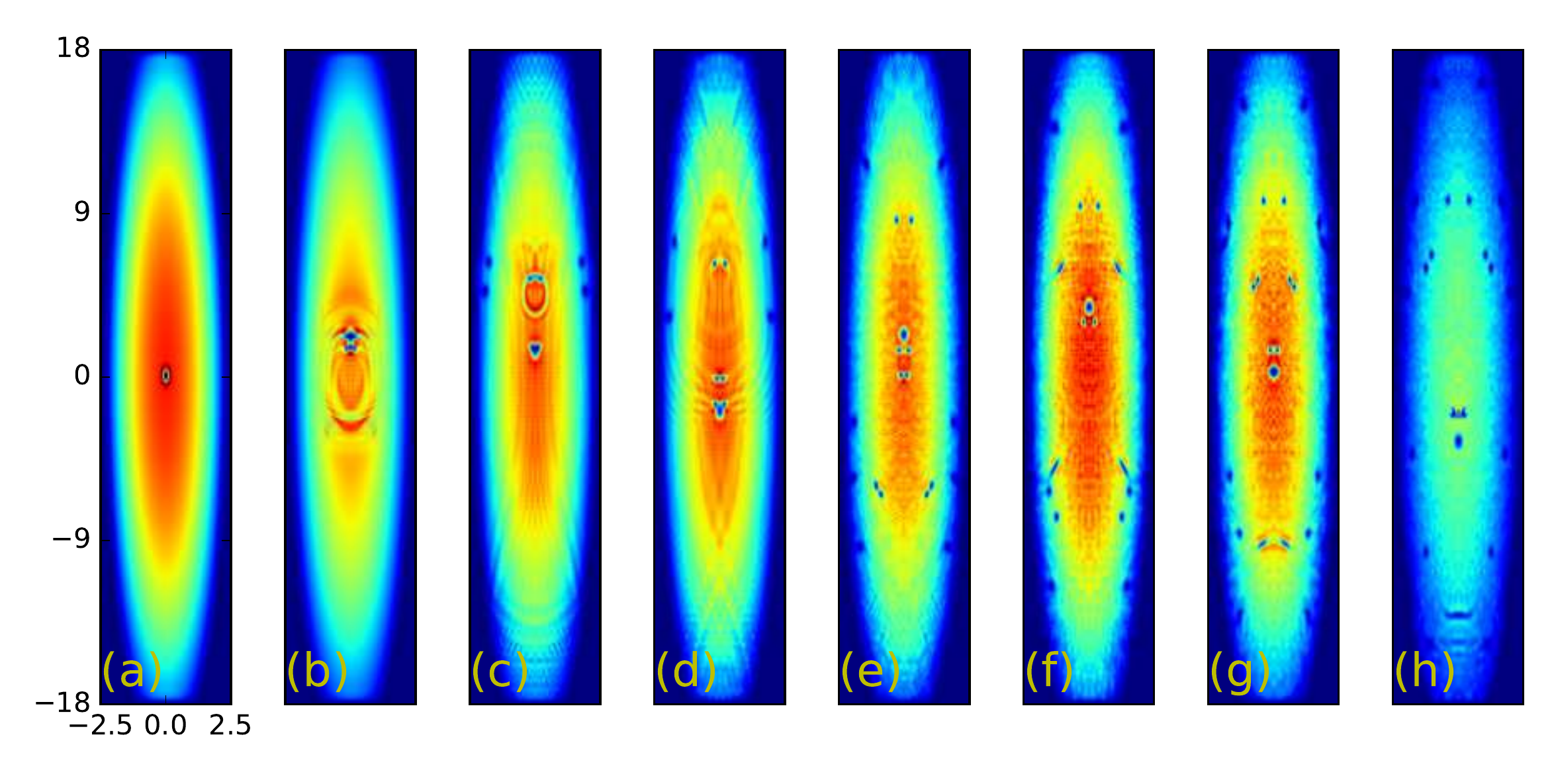}
\end{center}
\caption{(color online) Dynamics of the vortices for (upper panel) $^{168}$Er BEC ($a_{dd}=66a_0$) and (lower panel) $^{164}$Dy BEC ($a_{dd}=131a_0$) and $a_s=50a_0$ for all cases. From left to right $t=0\,$ms,  $t=0.4\,$ms, $t=2.0\,$ms, $t=2.6\,$ms, $t=4.9\,$ms, $t=5.9\,$ms, $t=6.7\,$ms, and $t=7.5\,$ms, respectively. }
\label{f2}
\end{figure}

First, we calculate the critical velocity for the nucleation of vortex dipoles with respect to scattering length as shown in in Fig.~(\ref{f1}) (a) for the three different cases, non-dipolar $a_{dd}=0\,a_0$ (Green line with circles), $^{168}$Er dipolar $a_{dd}=66\,a_0$ (Blue line with triangles), and $^{164}$Dy dipolar $a_{dd}=131\,a_0$ (Magenta line with stars) condensates, respectively. If, the velocity of the Gaussian potential ($V_p=\epsilon\,\, \omega$) is larger than the critical velocity $V_c$, vortex dipoles are nucleated in the condensate. As shown in Fig.~(\ref{f1}) (a), the critical velocity ($V_c$) for the nucleation of vortex dipoles increases with respect to increasing two-body contact and dipolar interaction strengths. This happens due to increasing interaction strengths leading to increase in the chemical potential of the condensate. Usually the height of the obstacle is determined with respect to the chemical potential. One needs to increase the height of the obstacle when the chemical potential increases. In this case, the critical velocity for nucleating vortices also depends on the height of the Gaussian obstacle. Also, the chemical potentials for the three different cases, alkali BEC (non-dipolar $a_{dd}=0$, $^{168}$Er dipolar BEC, and $^{164}$Dy dipolar BEC are 36.77, 44.65 and 64.55, respectively. Figure~(\ref{f1}) (a) shows that if we fix the amplitude of the obstacle then the critical velocity increases with respect to the interaction strengths. Usually, in rotating magnetic trap, for the nucleation of single vortex, the critical rotation frequency decreases with respect to increasing contact and dipolar interaction strengths. In the rotating trap, the vortex with same charge circulation is created whereas while stirring with an obstacle, we observe vortices with opposite charges circulations. Stirring beyond the critical velocity nucleates more vortices. This will be helpful to study the dynamics of multiple vortex dipoles. Another feature to look in Fig.~(\ref{f1}) (a) is the stability of the $^{164}$Dy dipolar BECs. We show the critical velocity of the $^{164}$Dy BEC from  $35\,a_0$, because below this scattering length the obstacle creates the local collapse. On the other hand, $^{168}$Er BEC is stable from $20\,a_0$. Figure~(\ref{f1}) (b) illustrates the plot of the $V_c/c_s$ Vs $a_s$ for the three different cases, nondipolar (Green line with circles)~\cite{ref1}, $^{168}$Er (Blue line with triangles), and $^{164}$Dy (Magenta line with stars) condensates, respectively. The speed of sound ($c_s$) depends on the scattering length and dipolar interaction strength ($c=\sqrt{2 n_0 \sqrt{2\pi} (a+a_{dd})/d_z}$). 

\begin{figure}[!ht]
\begin{center}
\includegraphics[width=0.7\linewidth]{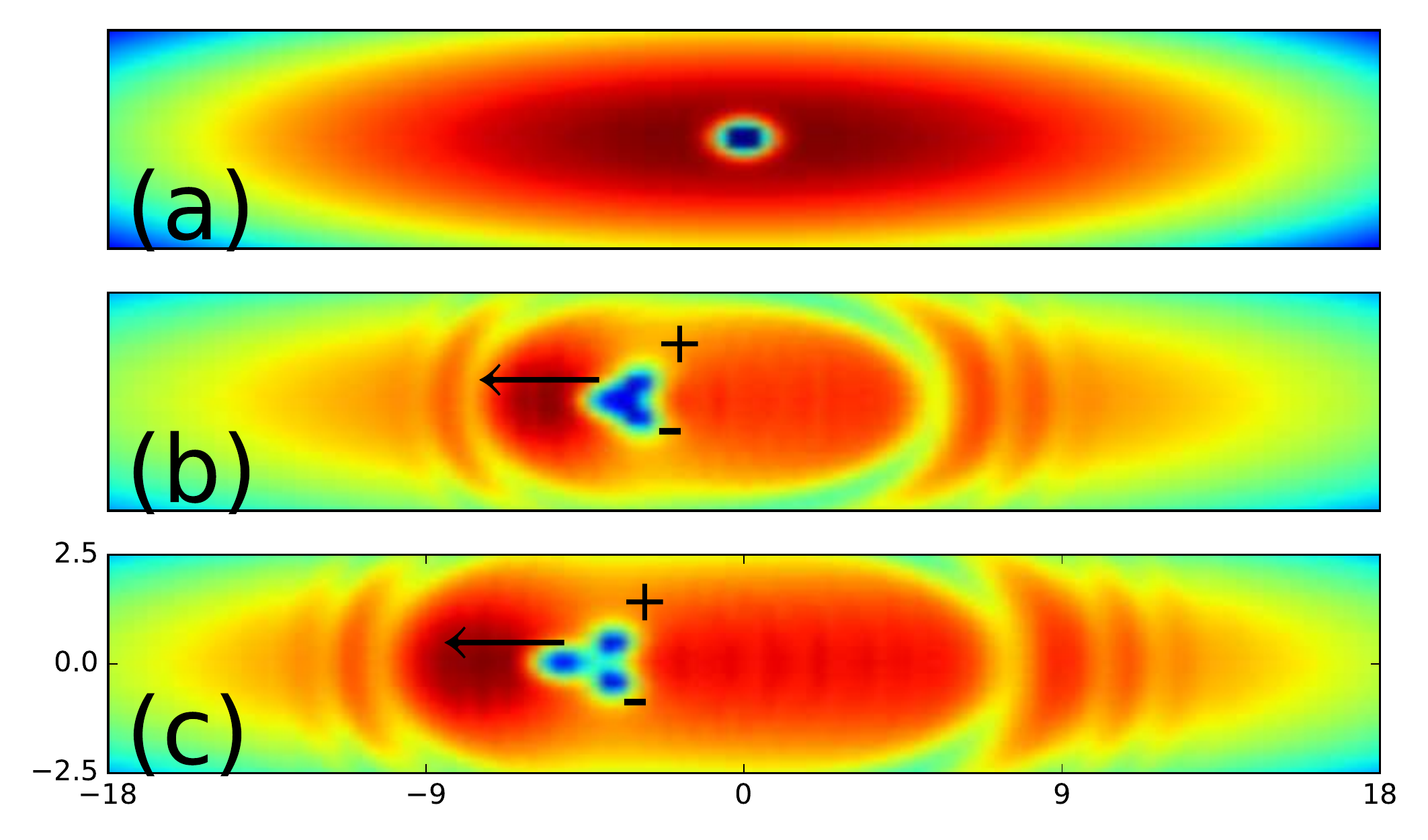}
\caption{(Color online) Density profile for nucleation of a vortex pair by the oscillating potential at (a) $t$ = 0 ms, (b) $t$ = 0.2 ms, and (c) $t$ = 0.4 ms are shown. The symbols $-$  and $+$ denote a vortex with clockwise or counterclockwise circulation, respectively. The black arrows indicate the direction of motion of the potential. A ghost vortex pair nucleates inside the potential (a), exits it (b), and finally fully leaves the potential  (c).}
\label{f3}
\end{center}
\end{figure}

The dynamics of vortices in which they experience a lengthy migration are shown in Figs.~(\ref{f2}) for two different cases. Following the destiny of vortices nucleated by the oscillating potential enables us to survey their dynamics. The initial state in the static Gaussian potential in Fig.~(\ref{f2}) is obtained by an imaginary time step of the GP equation. A vortex pair is nucleated behind the Gaussian potential in Fig.\hspace{0.5mm}~(\ref{f2})(b) as the potential starts to move. Then the oscillating potential nucleates vortex pairs whose impulses alternately change direction. They reconnect with each other to make new vortex pairs, leaving the potential in two cases. This phenomenon is not observed for the case of uniform motion of the potential, but only for an oscillating potential. Reaching the surfaces, the vortex pairs interact with ghost vortices, which are vortices in the low-density region. Then the vortices head toward the bow of the condensate along the surfaces. A vortex coming up from the left side reaches the bow to meet one from the right side, thus making a new vortex pair. Finally, the pair comes back to the center of the condensates. Thus the vortices nucleated by the potential enjoy a lengthy migration in the "sea" of BEC; the vortices are nucleated from the potential, reconnect, move away from it, reach the surface, head toward the bow and come back to the center. In the following, we illustrate elementary processes related to the synergy dynamics.

\begin{figure}[!ht]
\begin{center}
\includegraphics[width=0.7\linewidth]{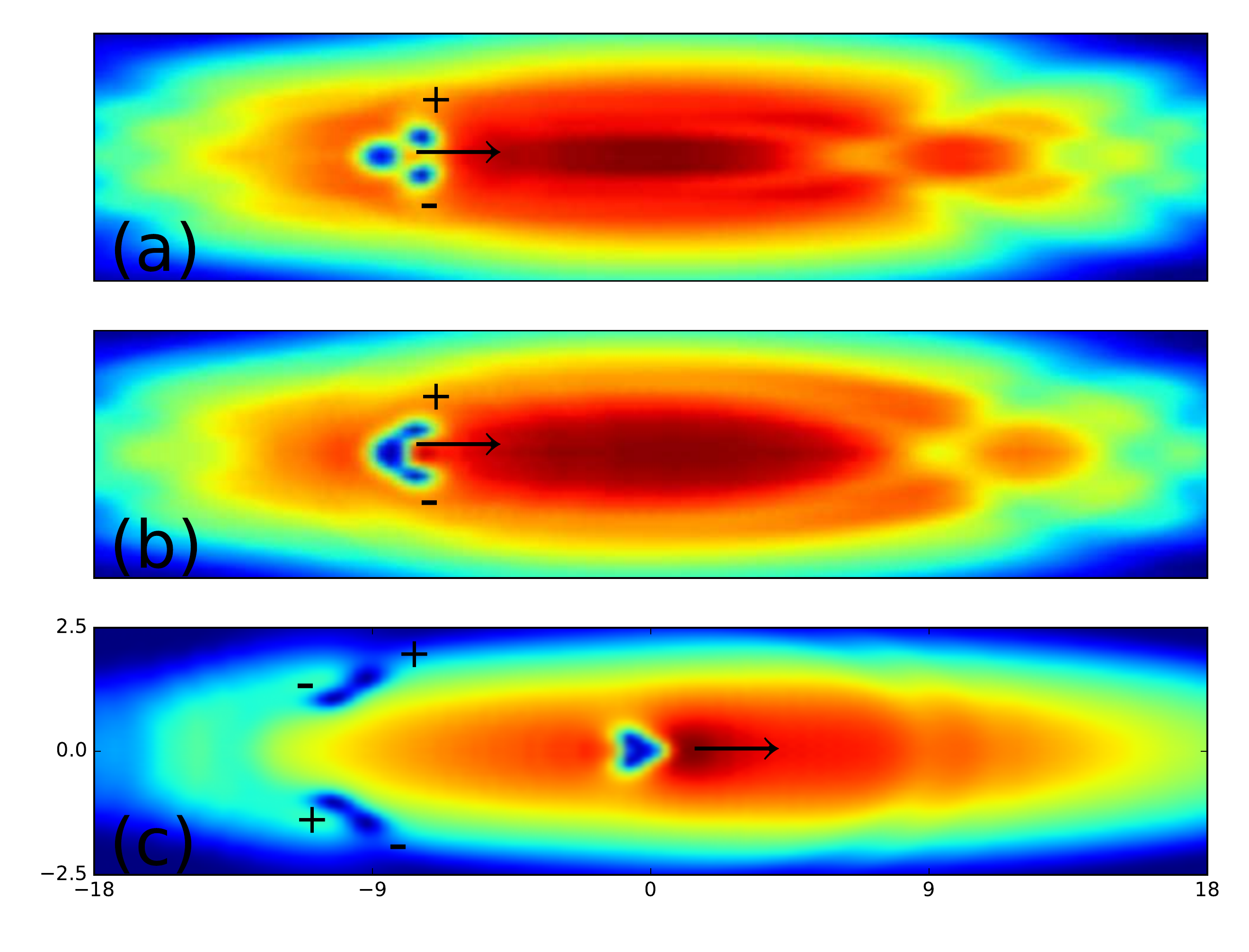}
\caption{(Color online) Density profile for reconnection of vortex pairs near the oscillating potential at (a) $t$ = 0.8 ms, (b) $t$ = 1.2 ms, and (c) $t$ = 2.3 ms are shown. The density profile before the collision between the potential and the vortex pair are shown in (a). Thereafter, another ghost vortex pair nucleates in (b), exiting the potential through the collision, which causes reconnection of the vortices. As a result, two pairs appear in (c).}
\label{f4}
\end{center}
\end{figure}
An oscillating potential creates vortex pairs, causes reconnection of pairs characterized by the oscillation, and causes the new pairs to leave for the surface of the condensate. Consequently, the surface becomes filled with vortices having positive and negative circulation, which leads to nucleation of rarefaction pulses and the migration of vortices. We call this sequence synergy dynamics of vortices and rarefaction pulses, which often occurs in cases where the amplitude of the oscillation is larger than the size of the oscillating potential. Ghost vortices, namely quantized vortices in a low density region, are important for the nucleation of the usual vortices in the bulk density region since nucleation requires seeds of vortices. In rotating BECs, ghost vortices are nucleated outside the condensate, entering it through the excitation of the surface waves, leading to the creation of usual vortices \cite{Tsubota02,Kasamatsu03}. Thus, the periphery of the condensate provides seeds of topological defects. In our system, the oscillating potential provides seeds within itself. The potential starts to move, inducing a velocity field like back-flow, emitting phonons, and a ghost vortex pair is nucleated inside the potential as shown in Fig.~(\ref{f3})(a). The ghost pair tends to move away from the potential in Fig.~(\ref{f3})(b), and a usual vortex pair appears in the condensate in Fig.~(\ref{f3})(c). Thus, the ghost vortices work as seeds of usual vortices. 

Reconnection of vortex pairs occurs near the oscillating potential. The new vortex pair has an impulse in the same direction as that of the potential. Then, the potential changes the direction of the velocity. Thus, the potential will collide with the pair in Fig.~(\ref{f4})(a). Then, a new ghost vortex pair is nucleated inside the potential whose impulse is opposite to that of the usual vortex pair, reconnecting with it as shown in Fig.~(\ref{f4})(b). Thus, two new vortex pairs appear in the condensate in Fig.~(\ref{f4})(c). Thereafter, the pairs move away from the potential, leaving for the surface of the condensate. This reconnection is characteristic of the oscillating potential because the potential repeatedly emits vortices of positive and negative circulation in opposite directions, which is not seen for potentials of uniform motion. This leads to nucleation of rarefaction pulses, as shown in the following.

\begin{figure}[!ht]
\begin{center}
\includegraphics[width=9cm,height=4.5cm]{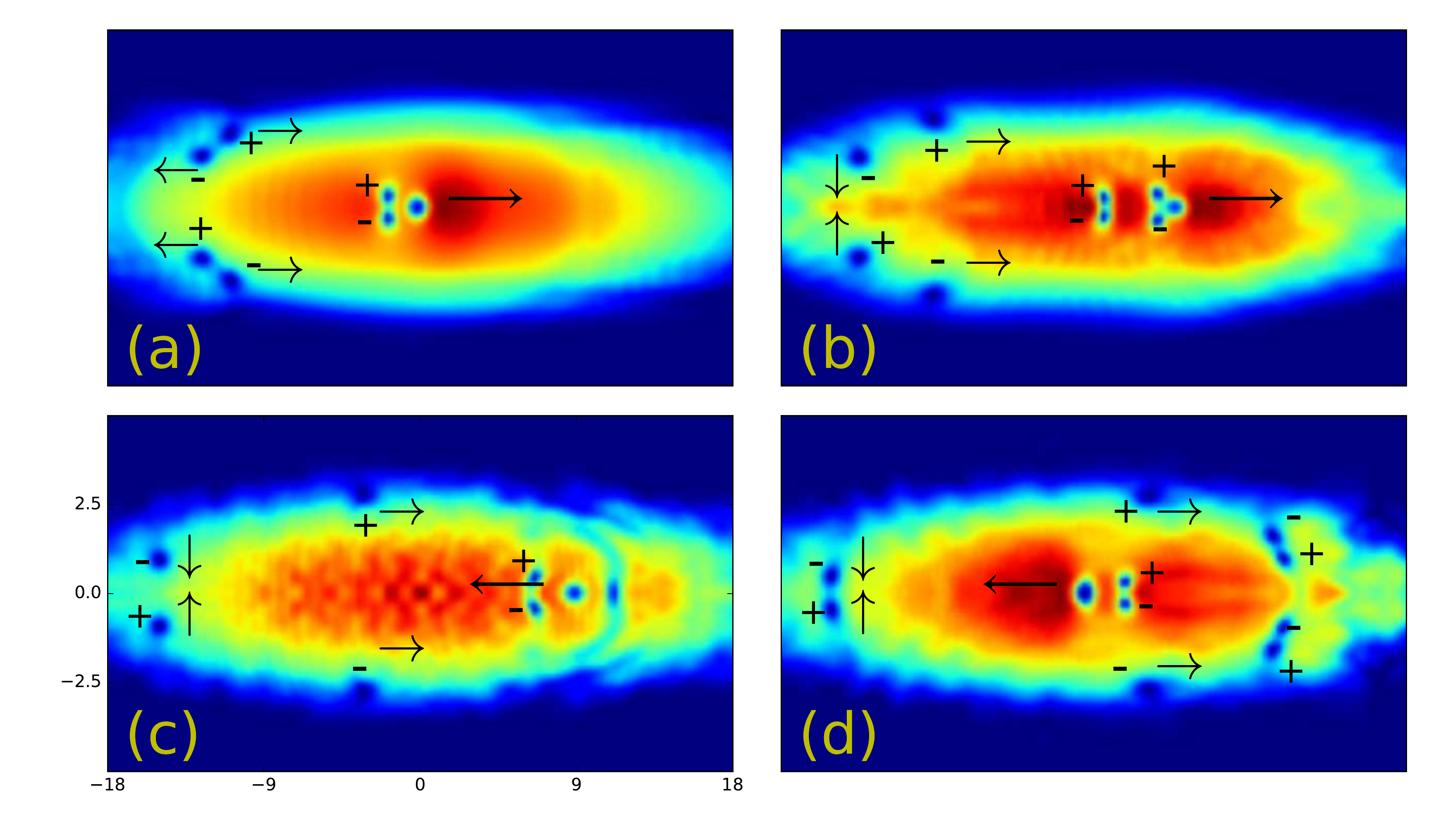}
\caption{(Color online) Density profile for nucleation of rarefaction pulses at (a) $t$ = 2.5 ms, (b) $t$ = 3.2 ms, (c) $t$ = 3.8 ms and (d) $t$ = 4.7 ms are shown. Some vortices sit near the surface in (a) and reconnection of the vortices occurs in (b). While the new pairs move toward the center of the condensate, the pairs annihilate, which leads to nucleation of rarefaction pulses in (c) and (d).}
\label{f5}
\end{center}
\end{figure}

The vortex pairs separate as they approach the surface of the condensate shown in Fig. (\ref{f5}). This behavior is qualitatively understood by applying the idea of an image vortex, which is often used in dynamics. The vortices induce a circular velocity field in a uniform system, but the field is distorted in a nonuniform system. This effect is strongly evident near the surface of the condensate where the density profile rapidly varies. As the vortices arrive at the surface, the normal component of the velocity field is suppressed. This situation is approximately equal to the relation between a vortex and a solid wall, so that the dynamics of vortices near the surface in Fig. (\ref{f5}) can be shown by the image vortex. Note that this idea only gives a qualitative understanding since the surface is not exactly a solid wall.  
The vortices near the surface of the condensate have two fates. One is that a vortex pair transforms into rarefaction pulses through the annihilation of the pairs. The other is that the vortices migrate in the condensate. We show these dynamics in the following. Many vortices have accumulated near the surface of the condensate in Fig.~(\ref{f5})(a) since the oscillating potential continues to make vortices with positive and negative circulation. Hence, the vortices near the surface can reconnect with each other as shown in Fig.~(\ref{f5})(b), where we enclose the new vortex pairs with square dotted lines. These pairs have impulse toward the center of the condensate. As a pair approaches the center, the 
size of the pair diminishes. Consequently, pair annihilation of vortices occurs, making the rarefaction pulses shown in Figs.~(\ref{f5})(c) and (d). 
The low density parts in Figs.~(\ref{f5})(c) and (d) have these properties and hence we can identify them as rarefaction pulses. This kind of nucleation of rarefaction pulses is characteristic of an oscillating potential since it is caused by the potential emitting vortices in opposite directions. 

\section{Conclusion}
\label{sec:con}
We have reported the dynamics of vortices in a dipolar Bose-Einstein condensate by solving the two-dimensional, nonlocal, Gross-Pitaevskii equation numerically. We have calculated the critical velocity for vortex nucleation and found that the critical velocity increases when we increase the strength of the dipolar interaction in the atomic BEC. We have showed the formation of the rarefaction pulses during the dynamics of the vortices with opposite rotations. We have reported the teaming dynamics of vortices and rarefaction pulses in a BEC with dipole-dipole interaction, which has not been reported upto now.

\vskip 0.5 cm
{\bf Acknowledgement}
The authors thank professors, P. Muruganandam, K. Porsezian, Bishwajyoti Dey and A. Gammal for their help. 
SS wishes to thank DST-SERB for offering a Post-Doc through National Post Doctoral Fellowship (NPDF) Scheme (Grant No. PDF/2016/004106). The work is partially supported by the UGC through Dr. D.S. Kothari Post Doctoral Fellowship Scheme (No.F.4-2/2006 (BSR)/PH/14-15/0046).
RKK acknowledges the financial support from FAPESP of Brazil (Contract number 2014/01668-8).

\end{document}